# Comment on "Superfluid density and specific heat within a self-consistent scheme for a two-band superconductor"

Annette Bussmann-Holder, Max-Planck-Institute for Solid State Research, Heisenbergstr. 1, D-70569 Stuttgart, Germany

Kogan et al. [Phys. Rev. B **80**, 014507 (2009)] present a two-band model to investigate its effect on superconducting properties as, e.g., the superfluid density, the specific heat, the dependence of the superconducting transition temperature $T_c$ on the model parameters. Due to a variety of simplifications and unphysical consequences of those, erroneous results and misleading conclusions are obtained.

In Ref. 1, Kogan et al. use a two-band model (TBM) to investigate various consequences of it for a multiband superconductor. The TBM has independently and simultaneously been introduced shortly after the BCS theory by Suhl et al. [2] and Moskalenko [3] in 1959 to account for a more complex Fermi surface topology than the one introduced by BCS. Extensions of this approach have been suggested shortly afterwards [4]. The important clue of these extensions to superconductivity is an interband pair scattering potential which leads to enhanced pair scattering via exchange through an additional channel. Early on it has been recognized that a $T_c$ enhancing effect is related to this extra channel.

In [1] a similar approach is used, however, introducing simplifications which lead to erroneous results and misleading conclusions.

The TBM in its simplest form reads:

$$H = H_0 + H_1 + H_2 + H_{12} \tag{1}$$

$$H_0 = \sum_{k_1\sigma} \xi_{k_1} c^+_{k_1\sigma} c_{k_1\sigma} + \sum_{k_2\sigma} \xi_{k_2} d^+_{k_2\sigma} d_{k_2\sigma} \tag{1a}$$

$$H_1 = -\sum_{k_1 k_1' q} V_1(k_1, k_1') c^+_{k_1+q/2\uparrow} c^+_{-k_1+q/2\downarrow} c_{-k_1'+q/2\downarrow} c_{k_1'+q/2\uparrow} \tag{1b}$$

$$H_2 = -\sum_{k_2 k_2' q} V_2(k_2, k_2') d^+_{k_2+q/2\uparrow} d^+_{-k_2+q/2\downarrow} d_{-k_2'+q/2\downarrow} d_{k_2'+q/2\uparrow} \tag{1c}$$

$$H_{12} = -\sum_{k_1 k_2 q} V_{12}(k_1, k_2) \{c^+_{k_1+q/2\uparrow} c^+_{-k_1+q/2\downarrow} d_{-k_2+q/2\downarrow} d_{k_2+q/2\uparrow} + h.c.\}. \tag{1d}$$

Here $\xi_{k_i}$ are the momentum $k$ dependent energies in band $i$, with creation and annihilation operators $c^+, c, d^+, d$; $V_i, V_{12}, V_{21}$ are the effective attractive interactions in band $i$=1, 2, and the interband interactions which mediate pairwise exchange between the two bands. Note, that the sign of the interband interaction is unimportant, opposite to the conclusions drawn in Ref. 3, since it always enters as square all properties of the superconductor.

By using standard techniques [5] coupled gap equations are derived from Eqs. 1:

$$\langle c^+_{k_1\uparrow} c^+_{-k_1\downarrow} \rangle = \frac{\Delta_{k_1}}{2E_{k_1}} \tanh\frac{\beta E_{k_1}}{2} = \Delta_{k_1} \Phi_{k_1} \tag{2a}$$

$$\langle d^+_{k_2\uparrow} d^+_{-k_2\downarrow} \rangle = \frac{\Delta_{k_2}}{2E_{k_2}} \tanh\frac{\beta E_{k_2}}{2} = \Delta_{k_2} \Phi_{k_2} \tag{2b}$$

$$\Delta_{k_1} = \sum_{k_1'} V_1(k_1, k_1') \Delta_{k_1'} \Phi_{k_1'} + \sum_{k_1} V_{1,2}(k_1, k_2) \Delta_{k_2} \Phi_{k_2} \qquad (2c)$$

$$\Delta_{k_2} = \sum_{k_2'} V_2(k_2, k_2') \Delta_{k_2'} \Phi_{k_2'} + \sum_{k_1} V_{2,1}(k_1, k_2) \Delta_{k_1} \Phi_{k_1} \qquad (2d)$$

This problem has to be solved simultaneously and self-consistently for each temperature T and defines $T_c$ when the solutions for both, $\Delta_{k_{ii}} = 0, i = 1,2$. In order to show the inadequacy of Ref. 1, similar notations and simplifications are introduced, namely, by converting the sums into integrals and ignoring any momentum dependence of the gaps, which means considering isotropic s-wave gaps, Eqs. 2c, d transform to

$$\Delta_{1_1} = \int_0^{\hbar\omega_D} N_1(0) V_{11} dE_{k1} \frac{\Delta_1}{\sqrt{E_{k1}^2 + \Delta_1^2}} \tanh \frac{\sqrt{E_{k_1}^2 + \Delta_1^2}}{2kT} + \int_0^{\hbar\omega_D} N_2(0) V_{12} dE_{k2} \frac{\Delta_2}{2\sqrt{E_{k1}^2 + \Delta_2^2}} \tanh \frac{\sqrt{E_{k2}^2 + \Delta_2^2}}{2kT}$$

(3a)

$$\Delta_2 = \int_0^{\hbar\omega_D} N_2(0) V_{22} dE_{k2} \frac{\Delta_2}{\sqrt{E_{k2}^2 + \Delta_2^2}} \tanh \frac{\sqrt{E_{k_1}^2 + \Delta_1^2}}{2kT} + \int_0^{\hbar\omega_D} N_1(0) V_{21} dE_{k1} \frac{\Delta_1}{2\sqrt{E_{k1}^2 + \Delta_1^2}} \tanh \frac{\sqrt{E_{k1}^2 + \Delta_1^2}}{2kT}$$

(3b)

where for simplicity and in analogy to [1] $\xi_k = E_k = E$. The assumption made in [1] that $(N_1 + N_2)/N(0) = 1$ restricts the system to a constant doping level and limits all further applications. Instead of using this constraint, the notations of [1] are further followed, even when ignoring important momentum dependencies in the band energies. Introducing $N_i(0)V_{ij} = \lambda_{ii,ij}$ the above Eqs. 3 further simplify to:

$$\Delta_{1_1} = \int_0^{\hbar\omega_D} \lambda_{11} dE_1 \frac{\Delta_1}{\sqrt{E_1^2 + \Delta_1^2}} \tanh \frac{\sqrt{E_1^2 + \Delta_1^2}}{2kT} + \int_0^{\hbar\omega_D} \lambda_{12} dE_2 \frac{\Delta_2}{2\sqrt{E_2^2 + \Delta_2^2}} \tanh \frac{\sqrt{E_2^2 + \Delta_2^2}}{2kT} \qquad (4a)$$

$$\Delta_2 = \int_0^{\hbar\omega_D} \lambda_{22} dE_2 \frac{\Delta_2}{\sqrt{E_2^2 + \Delta_2^2}} \tanh \frac{\sqrt{E_{2_1}^2 + \Delta_2^2}}{2kT} + \int_0^{\hbar\omega_D} \lambda_{21} dE_1 \frac{\Delta_1}{2\sqrt{E_1^2 + \Delta_1^2}} \tanh \frac{\sqrt{E_1^2 + \Delta_1^2}}{2kT} \qquad (4b)$$

In order to follow Ref. 1 another unwelcome simplification is used in Eqs. 4 by using a constant cutoff energy for all involved integrals, i.e., some Debye type frequency $\omega_D$. However, this has the advantage that now the integrals can be replaced by:

$$\lambda_{ii} \Delta_i \int_0^{\hbar\omega_D} \frac{dE_i}{\sqrt{E_i^2 + \Delta_i^2}} \tanh \frac{\sqrt{E_i^2 + \Delta_i^2}}{2kT} + \lambda_{ij} \Delta_j \int_0^{\hbar\omega_D} \frac{dE_j}{\sqrt{E_j^2 + \Delta_j^2}} \tanh \frac{\sqrt{E_j^2 + \Delta_j^2}}{2kT} = \lambda_{ii} \Delta_i F_i + \lambda_{ij} \Delta_j F_j$$

(5)

At this stage no more approximations are possible even if extreme cases like $\lambda_{ij} \ll \lambda_{ii}$ are considered since $F_1 \neq F_2$ for all T<$T_c$. If this were not the case $\Delta_1 = \Delta_2$ and a collapse to a single gap superconductor would result. The coupled equations impose a constraint on all involved coupling constants through the implicit equation:

$$\lambda_{12}^2 = \frac{(1-\lambda_{11}F_1)(1-\lambda_{22}F_2)}{F_1 F_2} \tag{6}$$

The above eq. 6 adopts in the case of $\lambda_{12} \neq \lambda_{21}$ the form:

$$\lambda_{12}\lambda_{21} = \frac{(1-\lambda_{11}F_1)(1-\lambda_{22}F_2)}{F_1 F_2} \tag{7}$$

Independent of which equation, 6 or 7, is used, $T_c$ is defined through

$$F_c = \int_0^{\hbar\omega_D} \frac{dE}{E} \tanh\frac{E}{kT_c} . \tag{8}$$

As is obvious from Eqs. 6 and 8 neither a sign change in $\lambda_{ij}$ nor in $\Delta_i$ as compared to $\Delta_j$ will affect the general conclusions. Especially, the results as depicted in Fig. 7 of Ref. 1 are not reproducible within the TBM, even if the above simplifications are introduced and eq.7 instead of eq. 6 is employed. Especially, it is obvious from eq. 7 that a sign change in either of the two $\lambda_{ij}$ corresponds to a negative right hand side of eq. 7 which is unphysical. In addition, the limit $\lambda_{ij} \to 0$ is only valid when T is extremely close to $T_c$ since this limit is analog to $1 = \lambda_{ii}\Delta_i F_i$ as can be easily seen from Eq. 6. As such, the results as shown in Fig. 4 of Ref. 1 are an artifact of the oversimplifications of the TBM. In order to demonstrate this, numerical calculations for the coupled gaps and the superfluid density $\rho_s(T)$, using the very simple form of Eq. 5 have been made (see Figs. 1a, b) where the interband coupling is of the order of $10^{-5}$ and the gap ratio is approximately the same as in Ref. 3. Obviously, a pronouncedly different T-dependence of both quantities is the result as compared to the results of Ref. 1. Especially it is noted that the low temperature inflection point in the superfluid density is shifted to temperatures in the vicinity of $T_c$.

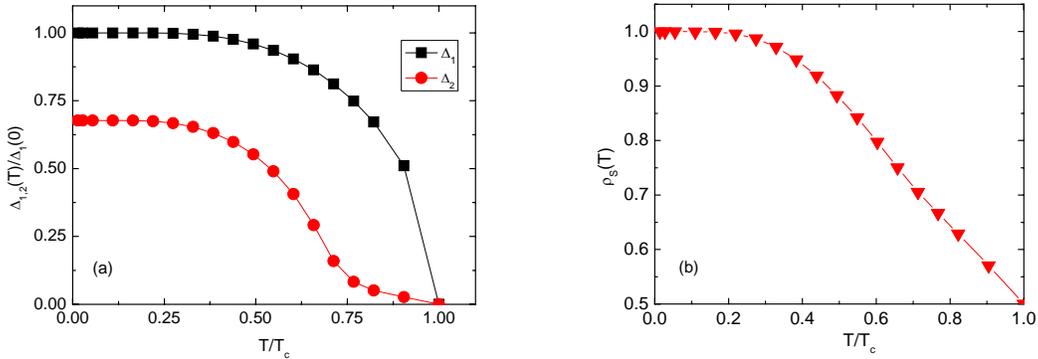

**Figure 1** Temperature dependence of the normalized coupled gaps $\Delta_{1,2}(T)/\Delta_{1,2}(0)$ with $\Delta_1$ being the larger of the two gaps (a), and the related superfluid density $\rho_s(T)$ (b).

To conclude, the model suggested in Ref. 1, even if employed in its most oversimplified version, leads to misleading results and erroneous conclusions and is inadequate to describe the physical properties of a TBM superconductor.